# Josephson-Like Behaviour of Granular Carbon Films

Sergey.G. Lebedev

*Institute for Nuclear Research of Russian Academy of Sciences*
*60th October Anniversary Prospect, 7a, Moscow, 117312, Russia*

**Abstract**

This work presents the study of some new anomalous electromagnetic effects in graphite-like thin carbon films. These are:

- The fast switching ($10^{-9}$ sec) of electrical conductivity
- The detection of microwave radiation and its temperature dependence
- The oscillations of film stack magnetization in the magnetic field of *1-5 T*.
- The optical radiation under process of spasmodic switching of conductivity

Results of magnetic force microscopy (*MFM*), *DC SQUID* magnetization, reversed Josephson effect (*RJE*), and resistance measurements in thin carbon arc (*CA*) films are presented. The observation of a *RJE* induced voltage as well as its rf frequency, input amplitude, and temperature dependence reveals the existence of Josephson-like Junction arrays. Oscillating behavior of the *DC SQUID* magnetization reminiscent of the Fraunhofer-like behavior of superconducting (*SC*) critical current in the range of *10000-50000 Oe* has been observed. The *DC SQUID* magnetization measurement indicates a possible elementary *102 nm* SC loop; this is compared to *MFM* direct observations of magnetic clusters with a median size of *165 nm*. The results obtained provides a basis for non-cryogenic elecrtonic devices utilizing the Josephson effect.

**Preconditions of carbon superconductivity**

The opportunity of existence of superconductivity above room temperature have substantiated by the Nobel prize winners V.L.Ginzburg [1] and W.A..Little [2].

After that many researchers with enthusiasm were started to searches of *HTSC*-materials.

According with the modern representations, superconductivity is caused by coupling of separate electrons with each others in Couper's pairs through a chain of atoms of a crystal lattice. Electrons as though constantly pull a chain, coordinating their movement with the partner. Thus the pair of electrons moves in a crystal lattice as a unity and does not dissipate its energy. The greater the exchanges frequency by "jerks", the more strongly the electrons are connected in pairs and the higher the temperature of destruction of a superconducting condition. It is noticed, that the «jerks frequency» is higher in the materials with a high melting point, such as carbon with its greater variety of chemical and structural forms. Therefore carbon and its compounds by one of the first «have got under suspicion». The known Polish scientist K.Antonowicz (*1914-2002*) more than *30* years ago investigated conductive properties of glassy carbon [3] and its raised dust deposits [4] and has found out the effect of jump of conductivity up to three orders of magnitude. Change of conductivity was reversable, and the relaxation time made some days. In the further Antonowicz has revealed increase in a current at an irradiation of an *Al-C-Al*-sandwich with the microwave irradiation [5]. However, change of a current occured with a time delay during *100 minutes*, after a microwave irradiation. At first sight such «time delay» are represented rather strange from the point of view of electronic mechanisms. Nevertheless, Antonowicz has explained the effect observed by the superconductivity at a room temperature [6]. The anomalous behaviour of carbon has been observed also by other resechers. G.M.Zhao and Y.S.Wang found the traces of *HTSC* with the critical temperature about *650°K* in the carbon nanotubes [7], Russian scientists V.I.Tsebro, O.E. Omelyanovskii and A.P.Moravskii observed the weak decay of *SC* currents in the composite made of carbon nanotubes at room temperature [8]. The fresh arguments in support of idea room temperature *SC* may be the resutls of N.Breda et al. about the possibility the *SC* with the $T_c \sim 320°K$ combined of fullerenes clusters $C_{28}$ [9] and also the work of K.K.Gomes et al. evidenced the existence of electrons pairing at temperatures well above of $T_c$ [10].

The author of given paper long time studied the application of thin carbon films as charge exchange targets for accelerators of ions. The physical model of failure of carbon targets under bombardement of ion beam has been developed. In agreement with the model developed the film failure explained by accumulation and anneling of mobile point radiating defects – the interstitials. Accumulation of these defects creates mechanical pressure in a thin carbon film which collapses when pressure will exceed ultimate strength of a foil material [11]. In the further by means of the developed model influence of a microstructure of a film and distinctions in conditions of its deposition on lifetime under an ion irradiation was investigated [12-14]. Results have been checked up experimentally on accelerators in the Laboratory of Nuclear Reactions of the Joint Institute for Nuclear Research (Dubna), Institute of Theoretical and Experimental Physics (Moscow), Institute of high Energy Physics (Serpukhov). During these works there was an interest to electromagnetic properties of carbon films.

**Strange electromagnetism**

Studying the conducting properties of carbon films, obtained by dispersion of graphite in an electronic arch, we have found out the jumps of electrical resistance on four-five orders of magnitude at some critical current [15] (see Fig.1). Then, with a delay, there came a relaxation of conductivity. All occured, as in Antonowicz experiments, but the results of its researches became known to us only many years later. At similarity of investigated effects and similarity of structure of samples the direction of our researches, their results and conclusions appreciably differ from that has been obtained by Antonowicz. At a room temperature the critical current varied within the limits of *5-500 mA* depending on type of a condensate and sample annealing conditions. With downturn of temperature the value of a critical current increased. After a relaxation low resistance state was completely restored, so samples can be used for switchings repeatedly. Time of switching of such contactless switch measured by us made *1* nanosecond that excludes the thermal mechanism of switching.

It is possible to explain a combination of so fast switching and long relaxation by the presence of Josephson's vortices opened in due time by Nobel prize winner A.A.Abrikosov working now in the Argonne National Laboratory, *USA*. These vortices represent the cylindrical objects limited by superconducting currents inside of which there is a kernel of a normal phase with the destroyed superconductivity. Each vortice bears in itself the one quantum of a magnetic flux. The vortices get into a film through boundary from the outside and can migrate under applied electric and magnetic fields, and also "to be hooked" for every possible defects and heterogeneity which always exist inside the film. Conditions of penetration depend on valuess of magnetic and electric fields. The greater the value of a magnetic field, the less the size of formed vortices and the easier they get and move in a film. The enclosed electric field "pushes out" vortices from a film. Therefore long relaxation time of a conductivity after switching can be connected with slow penetration of vortices in a film. At the same time the application of high enough electric field will neutralize the influence of fixing barriers and forces vortices to leave a film quickly. Actually the high conductivity of a carbon film is defined by movement of vortices under action of the enclosed voltage.

Other interesting feature of carbon films - occurrence of a constant voltage on contacts at an radiowave irradiation, i.e. detecting of the microwave radiation. Similar experiment in the elementary form also for the first time has been executed by Antonowicz. However we also knew nothing about it and consequently spent the experiment essentially differently and used technics and representations which were absent 30 years ago. Antonowicz considered a film as uniform Josphson's contact and observed changes in current-voltage characteristics of an *Al-C-Al*-sandwich under microwave radiation. Josephson's contact (*JC*) is formed between two superconductors divided by a thin layer of an insulator or normal metal. At currents below the critical value the Couper's pairs can tunneling from one superconductor to another, practically without destruction, and the *JC* behaves as a superconductor. Another words, if a current is below the critical value the voltage on contact is absent. But when the current reaches the critical value, the Couper's pairs collapse in a layer between two superconductors. Destruction of each pair is connected with emission of photon which frequency ν depends on electon coupling energy $E_b = \hbar v$, where $\hbar$ – is the Planck's constant. Such process refers to as non-stationary Josephson's effect and explains the emission of light radiation of a *JC*. It is knows also the reversed Josephson's effect – the inducing on a *JC* of a constant voltage under lighting. The reversed Josephson's effect is actively used at research both single *JC*, and their associations – Josephson's media (*JM*). 30 years ago representations about them have not been developed yet.

When the structure of carbon films has been studied, it became clear, that they represent conglomerates of graphite-like granules - nanoclusters incorporated in the "matrix" of amorphous carbon [16]. Hence, the neighbouring granules divided by the isolating layer of amorphous carbon, form a *JC*. The electric properties of similar granular film very much remind the behaviour of *JM*. Therefore we have initially assumed, that such film represent a *JM*. We saw the problem in proving its presence. Now the methodology of *JM* identification is developed well enough. Successes in this direction have been appreciably reached owing to studying of new high-temperature superconductors (*HTSC*) which, as it is known, represent the *JM*. It was found out, that at microwave irradiation of *JM* the constant voltage is induced, i.e. there is reversed Josephson's effect. This process reminds rectification of an alternating current, but is essentially distinct from the last.

To identify unequivocally the *JM*, it is necessary not only a technique of distinction of the reversed Josephson's effect and diodic rectification, but also rejection thermal emf and other side effects. Such technique has been developed by J.T.Chen with colleagues [17]. They investigated an astable impurity of a superconducting phase with critical temperature *240˚K*, containing in the sample of *HTSC* - ceramics with the critical temperature *90˚K*. The reaction of *JM* on the microwave radiation, the dependence on temperature, frequency and amplitude of the microwave signal has been as a result thoroughly studied. Owing to this technique it was possible to prove the presence of the *HTSC* - phase with $T_c = 240˚K$, that considerably exceeds the temperature limit reached for today of *130˚K*. We have applied the technique to research of a prospective *HTSC* - phase in carbon films. During experiments all of characteristic reactions of *JM* have been reproduced and therefore its existence in a carbon film is proved. The critical temperature of the *HTSC* - phase is defined as a point where the constant microwave induced voltage $V_{dc}$ tends to be zero. On the plot of dependence $V_{dc}$ vs temperature of a carbon film (Fig.2) it is possible to see, that $T_c$ makes *650˚K*.

The observable behaviour very much reminds "hot" superconductivity. However there is a question: why the carbon a film have a finite electrical resistance? The matter is that superconducting systems not always can get the general zero resistance or, speaking more precisely, a condition «the general phase coherence». It becomes possible, when resistance of a film in a normal state (e.g.at temperature above critical) less than characteristic value $RQ = 7$ *kOhm*. However as it can be seen from Fig.1, the normal state resistance of a carbon film makes tens *MOhm*. Apparently, the *SC*-phase borrows only small part of volume of the film sample that can explain its finite resistance. And whether there are bases to assume, what superconductivity is possible in nanosized graphite granules? Apparently, the answer is yes! V.L.Ginzburg has predicted the possibility of high-temperature superconductivity in the sandwiches made from hihgly conductive phase, surrounded by the dielectric with high dielectric permeability $\varepsilon$ [18, 19]. As it was mentioned, in a carbon film graphite granules are shipped in a matrix of amorphous carbon. By our estimations, dielectric permeability in graphite grains of granules makes *$\varepsilon = 15$* [20] (usually this value of the order of unit). And it allows to consider granulated carbon film as a direct embodiment of Ginzburg idea.

**Magnetic properties of carbon films**

Detecting of the microwave in the carbon film samples – is only one illustration of *JM* reality. Other evidences of Josephson's behaviour are shown in their magnetic properties. The measurements of magnetization of samples with a small fraction of a superconducting phase – is the very labour-consuming problem which can be solved only with the help of such high-sensitivity devices as *SQUID* (Superconducting Quantum Interference Device)-magnetometr. Action of this device is based on an interference of weak magnetic fluxes of the sample with a known magnetic flux in a superconducting ring in which the *JC* it is connected. In fact the Josephson's interference allows to measure the values of magnetic fluxes, comparable with quantum of magnetic flux *$\Phi_0 = 2 \cdot 10^{-7}\ Hauss \cdot cm^2$*. This value has dimension of the magnetic field by the area. If the area of a vortice makes *$1\ cm^2$* it bears a magnetic field *$2 \cdot 10^{-7}\ Hauss$*.

*SQUID*- magnetometr – is very dear device, therefore their number in the world is not so much. To us has had the luck to produce the similar measurements in the Laboratory of Superconductivity and Magnetism of the Leipzig University (Germany). In parallel with *SQUID* measurements a carbon film were studied in a Magnetic Force Microscope (*MFM*). The given device, also extremely expensive, is characterized by that it simultaneously allows to see both magnetic, and topological clusters in a film.

By means of *SQUID*-magnetometr we have found out the magnetization oscillations of the sample in the field of magnetic fields of $10^4$-$5 \cdot 10^4$ *Hauss* [21] (see Fig.3). The value of a magnetic field corresponding to jumps of magnetization, and also their amplitude depend on temperature (see Fig.4). Each oscillation is connected with the increase of a magnetic flux on one quantum $\Phi_0$ in the cluster. Using data of measurements, we have defined the average size of magnetic clusters - about *0.1 microns*. In the Magnetic Force Microscope we have seen the magnetic clusters and have defined their average size which has made *0.16 microns*. This good enough concurrence to the size found out in the *SQUID*- measurements. Having compared magnetic and optical "pictures", we have noticed, that, at least, some magnetic clusters coincide with topological ones (see Fig.5). Well distinguished currents on the picture flow round borders of clusters which remind the magnetic vortices. Whether the clusters visible in a *MFM* are the magnetic vortices? While it precisely is not known, check demands a new cycle of magnetic measurements.

**Our future plans**

In the further researches we assign greater hopes on doping of carbon films with the purpose of increase of their conductivity. Probably, it will allow to reach a zero resistance state at room temperature. Other perspective direction – is the search of optical radiation, possibly, emitting by a film during the moment of switching from a condition with high conductivity in a condition with high resistance. In fact if we deal with the *JM*, during the moment of destruction of superconductivity it is necessary to expect Josephson's radiation with the characteristic frequency defined by Couper's pairs coupling energy. For a superconductor with the critical temperature of *650 °K* it is necessary to search for infra-red radiation with the wavelength of some micrometers. The first attempts to register of such kind radiation by means of the fast high-sensitivity photodiode have appeared encouraging: it was possible to fix a series of optical impulses in the expected range, i.e. in a time vicinity near to spasmodic change of conductivity [22] (see Fig.6). The impulses amplitudes considerably exceed a level of "substrate" so enters the photo diode into a condition of saturation. However duration of impulses essentially exceeds expected value in *1* nanosecond. Apparently the existence of few impulses istead of one can be explained in therms of "multistage" generation due to gradual "deenergizing" of separate superconducting clusters. In that case switching also should be step-by-step, and the general duration of processes of switching and radiation can quite make the *100-th* fraction of second (see Fig.6). The spreadinf of an impulse of radiation relatively of initial nanosecond width of switching impulse should be connected with expenses of time on «brightening» and propagation processes of an optical impulse through a film and substrate body.

We will continue to investigate the *JM* in carbon films and simultaneously we reflect on possible applications about already obtained results. The existence of *JM* at room temperatures opens prospect of creation of various devices of noncryogenic Josephson's electronics [23].

One of similar application – is the contactless field effect switcher (see Fig.1). Integration of such switchers will essentially raise safety in electric networks and security from interferences. In the *USA* the similar field switcher on the basis of fullerenes [24] for a long time is under development, however it demands the cooling by liquid helium.

Other interesting application can be the Josephson's detector of γ-radiation for registration of neutrinos and a dark matter [25, 26].

It is possible to create the generators and detectors of the microwave radiation. However before it is necessary to analyse, whether they will have any advantages in comparison with existing devices.

It seems to be interesting to use the granular carbon films as magnetic protection. Influence of an electric current destroys magnetic vortices in a film. Returning in highly conductive state occurs at "pumping" of vorticess from the outside with "absorption" of magnetic fields from surrounding space. Facing by a carbon film of the walls, ceiling and a floor of rooms with increased requirements on magnetic protection would allow to supervise and prevent penetration of magnetic fields by measuring the electrical resistance of separate sites of a covering.

Optical radiation of films at spasmodic switching of conductivity opens one more opportunity - designing of new type of lasers on magnetic vortices. The power of the similar laser can appear rather significant. As show our experiments, during the moment of switching the power developed by a current, makes some Watts. As a result of jump the resistance of a film during 1 nanosecond increases in *10* thousand times, i.e. the film turns practically to an insulator. Where all power in some Watts of that direct current of electrons, which 1 nanosecond before rushed with huge speed, disappears? It reminds the crush-test of the car when with the purpose of check of a pillow of safety it on the full speed runs into a concrete wall. Apparently, all disappeared capacity brighted in the form of several impulses of radiation with the duration in 1 nanosecond each, and we obtain the pulse power of such kind laser on magnetic vortices of the order 1 GW.

The interesting application can appear a covering a carbon film of an internal surface of resonators for accelerators of elementary particles, and also systems of the microwave - transportations of the electric power with the purpose of reduction of losses, and creation of *HTSC* - wires for powerful superconducting noncryogenic magnets. The deposition technology of carbon film coverings easily allows to make it from a gaseous phase (*CVD*-technology). Moreover, this process without any problems can be automated, supervising quality of a covering through the measuring of their electroresistance.

All listed allows to hope, that for the granular carbon films the light future in power and electronics can be predicted.

**Acknowledgements**

I would like to thank Academicians of RAS V.A.Matveev and V.L.Ginzburg and Profs. V.L.Kravchuk A.I.Golovashkin for support of this work. I also acknowledge the support from the Russian Foundation of Basic Researches through the grants 05-08-17909-a, 06-08-99003-c, and 07-08-08099-з.

**References**

1. Ginzburg V.L. // Sov.Phys., ZhETP. 1964. T.47. C.2318-2327.
2. Little W.A. // Phys. Rev. 1964. V.134. P.A1416-A1424.
3. Antonowicz K. et al. // Carbon. 1973. V.11. P.1-5.
4. Antonowicz K. et al. // Carbon. 1972. V.10. P.81-86.
5. Antonowicz K. // Phys. Status. Solidi (a.) 1975. V.28. №2. P.497-502.
6. Antonowicz K. // Nature. 1974. V.247. P.358-360.
7. G.M.Zhao, Y.S.Wang.// ArXiv: cond-mat/0111268v2
8. V.I.Tserbo, O.E.Omelyanovskii and A.P.Moravskii.// JETP Lett., 70(1999)462.
9. Breda N., Broglia R.A., Colo G., Onida G., Provasi D., and Vigezzi E. // Phys.Rev.B62(2000)130-133.

10. Gomes K.K., Pasapathy A.N., Aakash Pushp, Shimpei Ono, Yoshi Ando, and Ali Yazdani.// Nature, 447(2007)569.
11. Koptelov E.A., Lebedev S.G., Panchenko V.N. // Nucl. Instr. Meth. 1987. V.A256. P.247-250.
12. Koptelov E.A., Lebedev S.G., Panchenko V.N. // Nucl. Instr. Meth. 1989. V.B42. P.239-244.
13. Lebedev S.G. // Nucl. Instr. Meth. 1994. V.B85. P.276-280.
14. Lebedev S.G. // Nucl. Instr. Meth. 1995. V.A362. P.160-162.
15. Lebedev S.G., Topalov S.V. // Bulletin of Lebedev's Physical Institute, N12(1994)14-20.
16. Lopinski G.P. et al. // Phys. Rev. Lett. 1998. V.80. P.4241-4250.
17. Chen J.T. et al. // Phys. Rev. Lett. 1987. V.58. P.1972-1975.
18. Ginzburg V.L. // Uspekhi Fizicheskikh Nauk, V.1968. T.95. C.91-99.
19. Ginzburg V.L. // Uspekhi Fizicheskikh Nauk, 1991. T.169. C.25-34.
20. Lebedev S.G. // ArXiv:cond-mat/0510304
21. Lebedev S.G. // Nucl. Instr. Meth. 2004. V.A521. P.22-29.
22. Lebedev S.G., Lebedev A.S., Yants V.E. // ArXiv: cond-mat/0612451.
23. Lebedev S.G. // Patent of the Russian Federation №2212735, 2003.
24. Schon J.H., Kloc Ch., Haddon R.C., Batlogg B. // Science. 2000. V.288. P.656-659.
25. Da Silva A. et al. // Proc. Of Workshop on Low Temperature Detectors for Neutrinos and Dark II. France, 1988. P.417-432.
26. Beck C., Mackey M.C. // ArXiv:astro-ph/0406504.

Figures captions

**Fig.1**. The current-voltage characteristic of the electronic contactless switcher at two variants of switching - manual (squares) and programing (cercles) adjustment of a current on the sample. Time of switching makes *1* nanosecond.

It is visible, that electroresistance of the sample during switching increases for four-to-five orders of magnitude.

**Fig.2.** The dependence of amplitude of the induced *DC* voltage under microwave irradiation vs sample temperature. The amplitude goes to zero at temperature *650˚K*.

**Fig.3.** The magnetization oscillations of the sample depending on the enclosed magnetic field for different temperatures *10*, *100*, *200* and *300˚K*. The form of the curve is caused by action of three factors: the primary "hill" is caused by the presence of ferromagnetic impurities, then due to natural diamagnetism of graphite magnetization linearly decreases, and on this background the spasmodic oscillations arise, which are connected with the entering of quantums of a magnetic flux in the magnetic clusters.

**Fig.4**. The ocsillations area «under a microscope». It is visible, that oscillation peaks are shifted down to lower magnetic field, and their amplitude decrease with the growth of temperature.

**Fig.5.** The simultaneous image of a surface of a carbon film in optical (at the left) and a magnetic microscope (on the right). Correlations in position, of at least, some magnetic and topological clusters are visible. Along borders of the magnetic clusters the circulating currents are seen. On a photo this currents look like as dark strips along a cluster's chain and remind huge boulders in the mountain river if to look onto it from the height of the bird's flight.

**Fig.6.** The generation of optical radiation in a time vicinity near to spasmodic change of conductivity. It is visible, that generation occurs step-by-step during several impulses which amplitudes decrease with the completion of process.

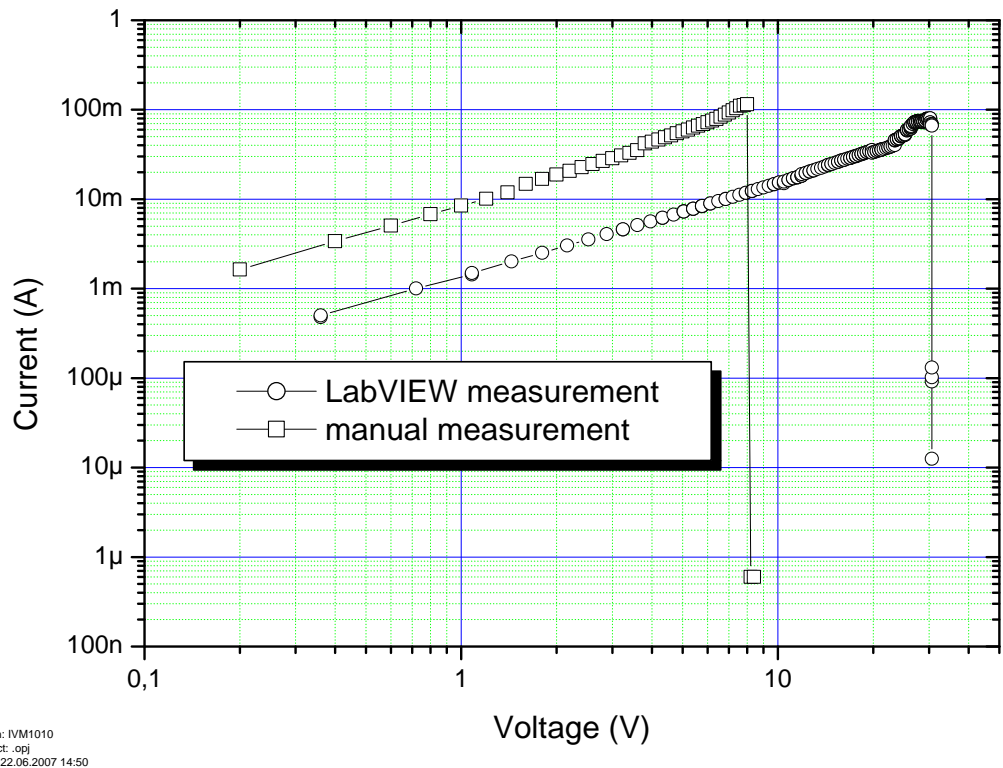

Fig.1.

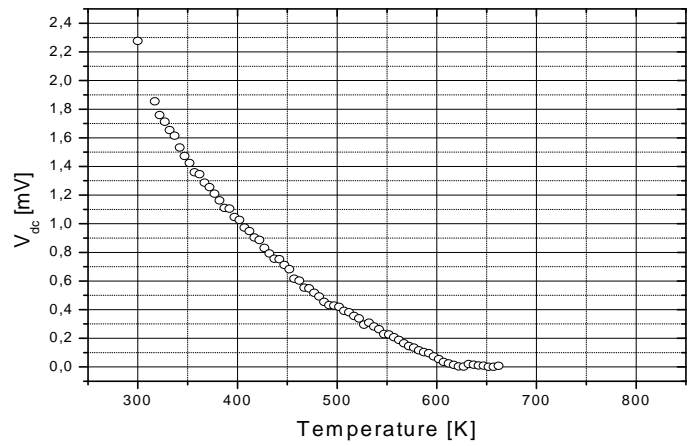

Fig.2.

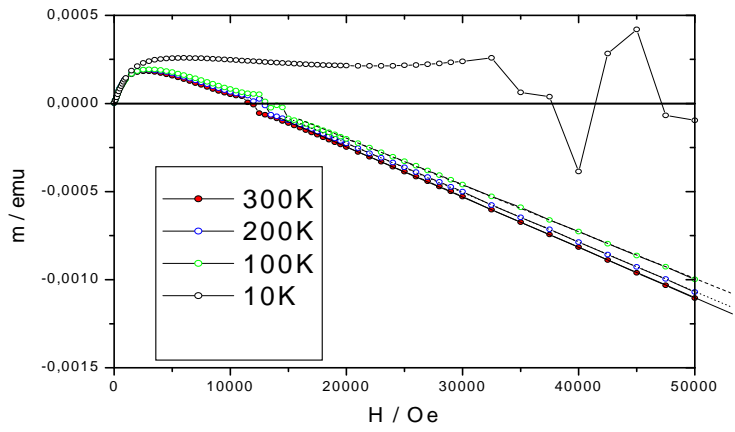

Fig.3

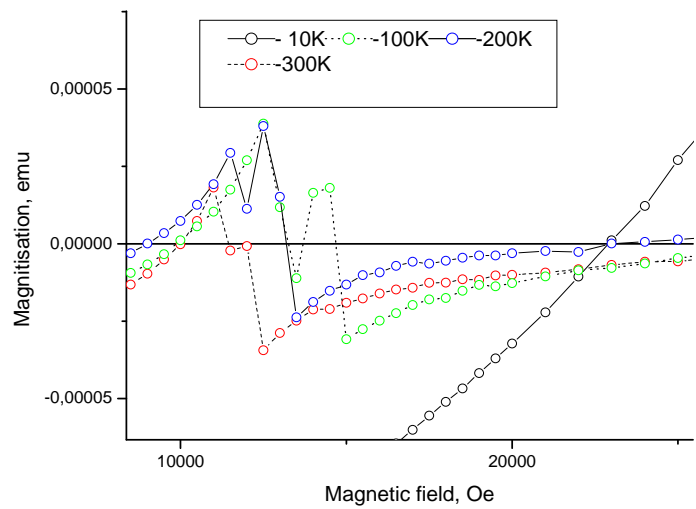

Fig.4.

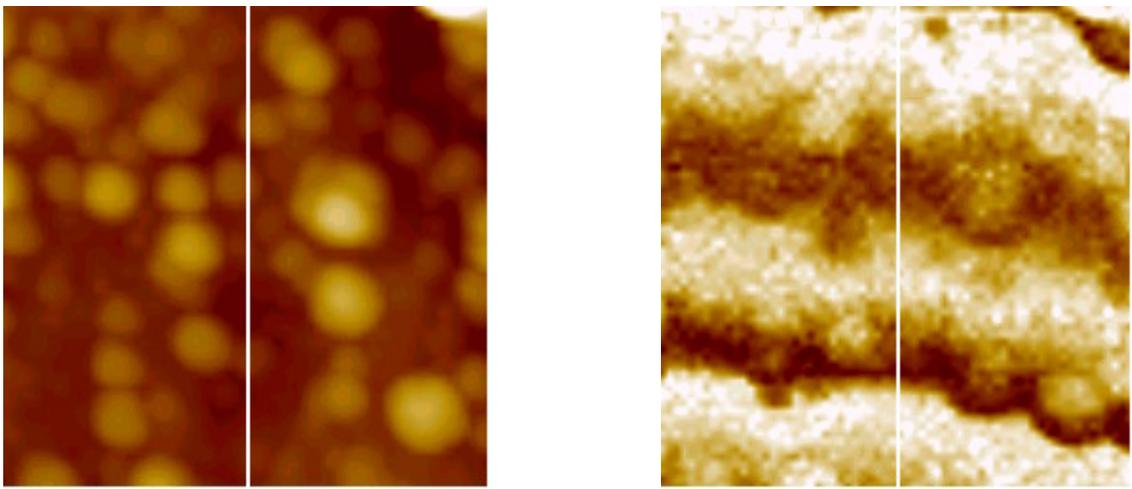

Fig.5.

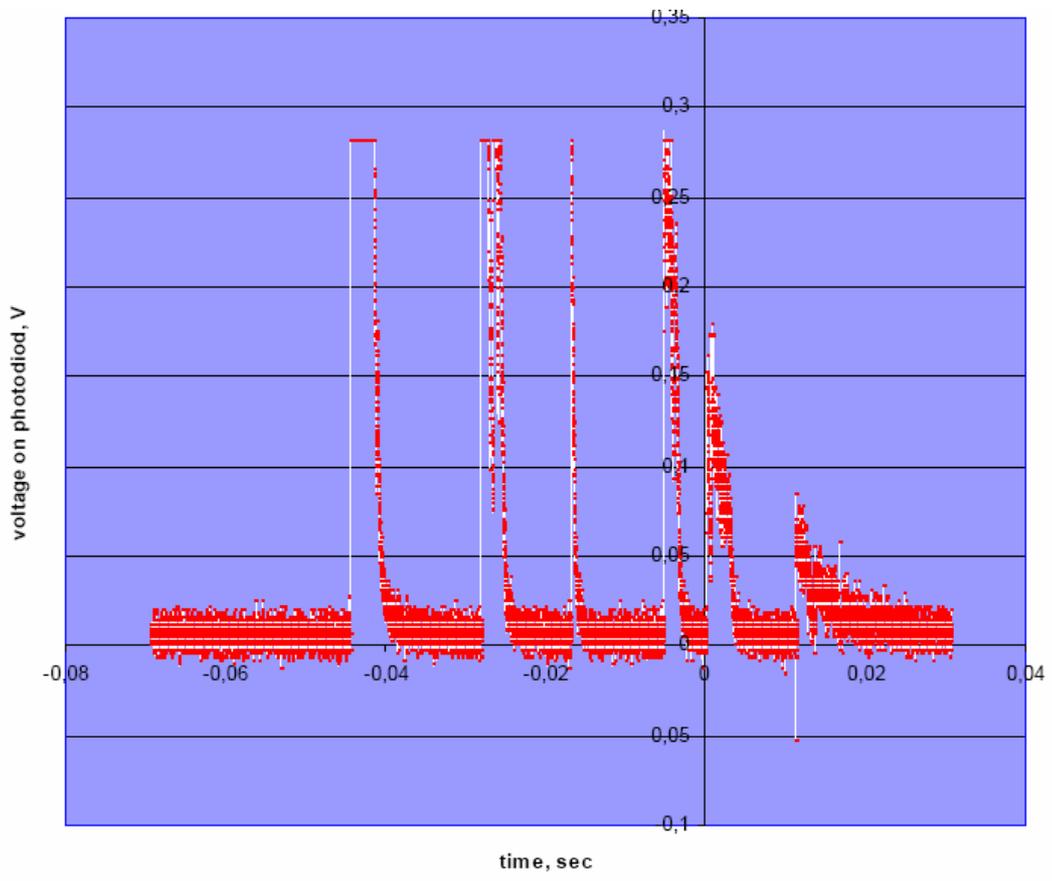

Fig.6.

This document was created with Win2PDF available at http://www.daneprairie.com.
The unregistered version of Win2PDF is for evaluation or non-commercial use only.